\documentclass{elsart1p}
\usepackage{amssymb}
\usepackage{graphicx}
\usepackage{epsfig}
\begin{document}
\begin{frontmatter}
\title{The Giant Monopole Resonance in the $^{112-124}$Sn Isotopes and
the Symmetry Energy Term in Nuclear Incompressibility  }

\author[label1]{T. Li}
\author[label1]{U. Garg}
\author[label1]{Y. Liu}
\author[label1]{R. Marks}
\author[label1]{B. K. Nayak}
\author[label1]{P. V. Madhusudhana Rao}
\author[label2]{M. Fujiwara}
\author[label2]{H. Hashimoto}
\author[label2]{K. Kawase}
\author[label2]{Y. Nakanishi}
\author[label2]{S. Okumura}
\author[label2]{M. Yosoi}
\author[label3]{M. Itoh}
\author[label3]{M. Ichikawa}
\author[label3]{R. Matsuo}
\author[label3]{T. Terazono}
\author[label4]{M. Uchida}
\author[label5]{T. Kawabata}
\author[label6]{H. Akimune}
\author[label7]{Y. Iwao}
\author[label7]{T. Murakami}
\author[label7]{H. Sakaguchi}
\author[label7]{S. Terashima}
\author[label7]{Y. Yasuda}
\author[label7]{J. Zenihiro}
\author[label8]{M. N. Harakeh}

\address[label1]{Department of Physics, University of Notre Dame, Notre
Dame, IN 46556, USA}
\address[label2]{Research Center for Nuclear Physics, Osaka University,
Osaka 567-0047, Japan}
\address[label3]{Cyclotron and Radioisotope Center, Tohuku University,
Sendai 980-8578, Japan}
\address[label4]{Department of Physics, Tokyo Institute of Technology,
Tokyo 152-8850, Japan}
\address[label5]{Center for Nuclear Study, University of Tokyo,
Tokyo 113-0033, Japan}
\address[label6]{Department of Physics, Konan University, Kobe 658-8501,
Japan}
\address[label7]{Department of Physics, Kyoto University, Kyoto 606-8502,
Japan}
\address[label8]{KVI, 9747 AA Groningen, The Netherlands}

\begin{abstract}
We have investigated the isoscalar giant monopole resonance (GMR) in
the Sn isotopes, using inelastic scattering of 400-MeV
$\alpha$-particles at extremely forward angles, including 0$^\circ$.
A value of $-550 \pm 100$ MeV has been obtained for the asymmetry
term, $K_\tau$, in the nuclear incompressibility.
\end{abstract}


\end{frontmatter}

Incompressibility of nuclear matter remains a focus of experimental
and theoretical investigations because of its fundamental importance
in defining the equation of state (EOS) for nuclear matter. The
compressional-mode giant resonances -- the Giant Monopole Resonance
(GMR) and the isoscalar giant dipole resonance (ISGDR), an exotic
compressional mode of nuclear oscillation -- provide a direct means
to experimentally determine the nuclear incompressibility. From
recent measurements on the GMR and the ISGDR, a value of
$K_{\infty}$ = 240 $\pm$ 10 MeV has been obtained,
%; this value is
consistent with results of recent theoretical results in both relativistic
and non-relativistic frameworks \cite{colo,jorge,shlomo}.

The asymmetry term, $K_{\tau}$, in the expression for nuclear
incompressibility is important because it is critical to determining
the radii of neutron stars, and also in understanding the
compressional-mode resonances in very neutron-rich nuclei, the
primary focus of  investigations at current and forthcoming radioactive ion beam
facilities. In particular, it has been shown that the radius of a
1--1.5 $M_{\odot}$ neutron star is mostly determined by the density
dependence of the asymmetry-energy term \cite{latt,jorge2}.

\begin{figure}[h]
\centering
\begin{tabular}{cc}
\epsfig{file=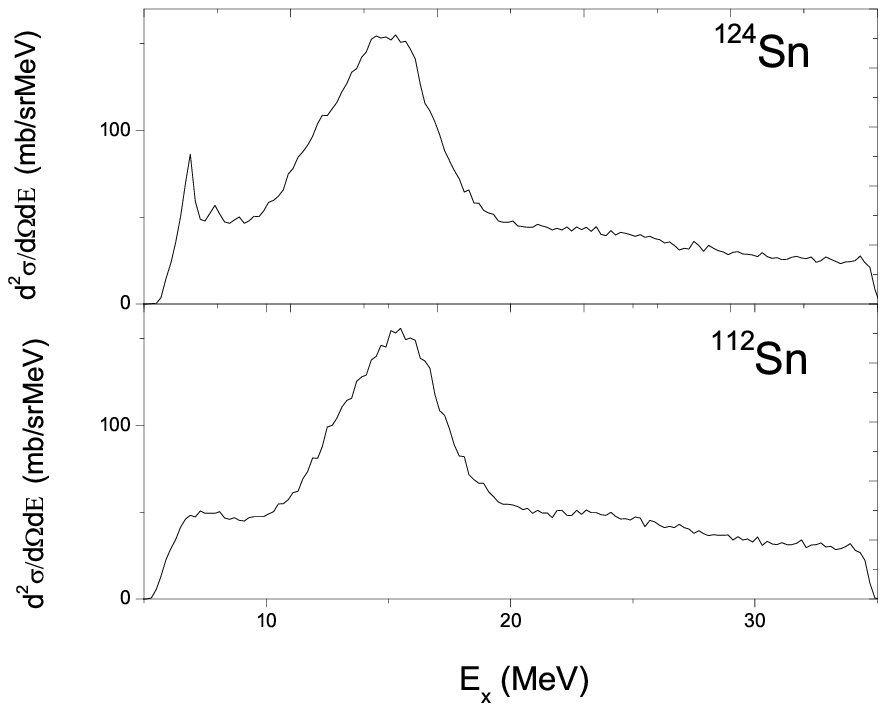,width=0.5\linewidth,clip=} &
\epsfig{file=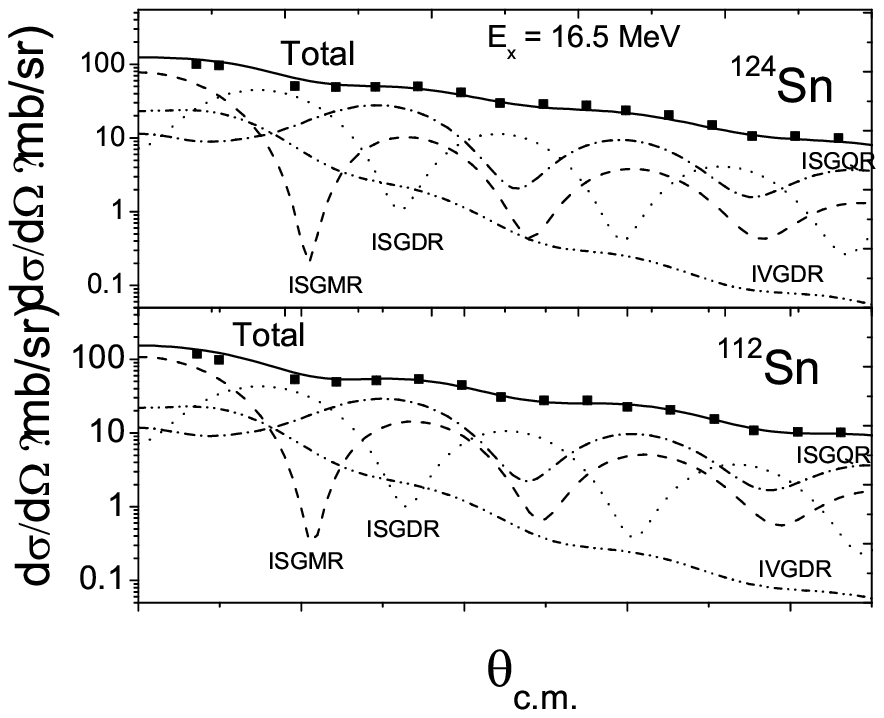,width=0.6\linewidth,clip=}\\
\end{tabular}

\vspace*{-0.8cm} \caption{\label{fig:ang} (a) 0$^{\circ}$ inelastic
scattering spectrum for $^{112}$Sn($\alpha$,$\alpha\prime$) and
$^{124}$Sn($\alpha$,$\alpha\prime$) at E$_\alpha$= 400 MeV.
(b) Angular distribution of 1-MeV bins centered at E$_x$=16.5 MeV for
$^{112}$Sn($\alpha$,$\alpha\prime$) and
$^{124}$Sn($\alpha$,$\alpha\prime$). The
solid squares are the experimental data and the solid line is the
MDA fit to the data. Also shown are the contributions to the fit from L=0
(dashed line), L=1 (dotted line), L=2 (dash-dotted line) and IVGDR
(dash-dot-dotted line).}
\end{figure}

In this presentation, we report on new measurements on GMR in the
even-A Sn isotopes. The experiment was performed at the ring
cyclotron facility of the Research Center for Nuclear Physics
(RCNP), Osaka University, using inelastic scattering of 400-MeV
$\alpha$ particles at extremely forward angles, including $0^\circ$.
Details of the experimental techniques and the data analysis
procedures have been provided previously \cite{uchida}.
Sample background-free ``$0^\circ$'' inelastic spectra are presented
in Fig.~\ref{fig:ang} (a). Giant monopole resonance strength
distributions were obtained for all Sn isotopes under study using a
 multipole-decomposition analysis (MDA) \cite{uchida}; Fig.~\ref{fig:ang} (b)
shows a sampling of MDA fits to angular distribution data.
% to angular distributions of 1-MeV-wide energy bins from the
%inelastic scattering spectra, centered at E$_x$=16.5 MeV.
%From the
%MDA analysis, GMR strength distributions were obtained for all
%isotopes.
Table I lists the centroid energies and widths extracted
from Lorentzian-peak fits to the extracted GMR strength
distributions, as well as the ``standard'' moment-ratios typically
used in theoretical calculations.

\begin{table*}[h]
\begin{center}
%\begin{ruledtabular}
\caption{\label{tab:gmr}Lorentzian-fit parameters and
  moment-ratios for the GMR strength distributions in the Sn
  isotopes.}
%, as extracted from MDA analysis in this work.}
\begin{tabular}{cccccc}

\hline Target\ \ \ \ \ &$E_{GMR}$ (MeV)\ \ \ \ \ &$\Gamma$ (MeV)\ \
\ \ \ &$m_1/m_0$
(MeV)\ \ \ \ \ &$\sqrt{m_3/m_1}$ (MeV)&\ \ \ \ \ $\sqrt{m_1/m_{-1}}$
(MeV) \\
\hline \vspace*{-0.25cm}
$^{112}$Sn   &$16.1\pm0.1$&$4.0\pm0.4$
&$16.2\pm0.1$&$16.7\pm0.2$&$16.1\pm0.1$ \\
\vspace*{-0.20cm}
$^{114}$Sn   &$15.9\pm0.1$&$4.1\pm0.4$
&$16.1\pm0.1$&$16.5\pm0.2$&$15.9\pm0.1$ \\
\vspace*{-0.20cm}
$^{116}$Sn   &$15.8\pm0.1$&$4.1\pm0.3$
&$15.8\pm0.1$&$16.3\pm0.2$&$15.7\pm0.1$ \\
\vspace*{-0.20cm}
$^{118}$Sn   &$15.6\pm0.1$&$4.3\pm0.4$
&$15.8\pm0.1$&$16.3\pm0.1$&$15.6\pm0.1$ \\
\vspace*{-0.20cm}
$^{120}$Sn   &$15.4\pm0.2$&$4.9\pm0.5$
&$15.7\pm0.1$&$16.2\pm0.2$&$15.5\pm0.1$ \\
\vspace*{-0.20cm}
$^{122}$Sn   &$15.0\pm0.2$&$4.4\pm0.4$
&$15.4\pm0.1$&$15.9\pm0.2$&$15.2\pm0.1$ \\
\vspace*{-0.10cm}
$^{124}$Sn   &$14.8\pm0.2$&$4.5\pm0.5$
&$15.3\pm0.1$&$15.8\pm0.1$&$15.1\pm0.1$ \\
\hline
\end{tabular}

%\end{ruledtabular}
\end{center}
\end{table*}

The incompressibility of a nucleus, $K_{A}$, may be expressed as:

\begin{equation}
K_{A} \sim  K_{vol}(1 + cA^{-1/3}) + K_{\tau}((N - Z)/A)^{2} +
K_{Coul}Z^{2}A^{-4/3}
\end{equation}

\noindent
% Here, c $\approx$ -1, and $K_{Coul}$ is essentially model independent.
For a series of isotopes, the difference $K_A-K_{Coul}Z^2A^{4/3}$
may, thence, be approximated to have a quadratic relationship with the
asymmetry parameter, [(N-Z)/A]. Such a quadratic fit, using the
customary moment ratio $\sqrt{m_{1}/m_{-1}}$ for the energy of the
GMR in calculating the $K_{A}$, gives K$_\tau$ = -550 $\pm$ 100 MeV;
the quoted uncertainty includes the uncertainties in the values of
$K_{A}$ and $K_{Coul}$, as well as the statistical uncertainties
from the fitting procedure.

\begin{figure}[h]
\vspace*{-0.4cm}
\begin{center}

\includegraphics[width=8cm]{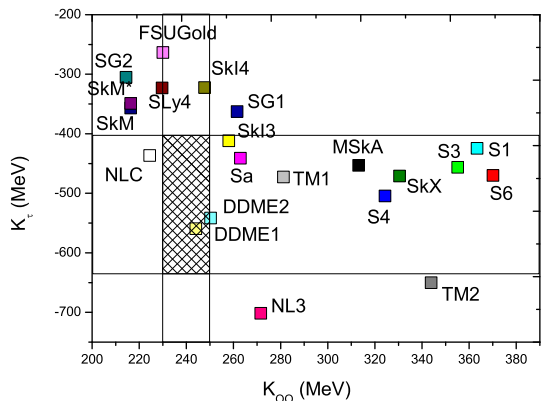}% Here is how to import EPS art

\vspace*{-0.8cm} \caption{\label{fig:kcomp}
Values of K$_\infty$ and K$_\tau$ calculated from
the parameter sets of various interactions as labeled \cite{hs}.
The vertical and horizontal lines indicate the experimental ranges of
K$_\infty$ and K$_\tau$, as determined from the GMR work.}
\end{center}
\end{figure}

From the data on the compressional-mode
giant resonances, we now have ``experimental'' values of both
K$_\infty$ and K$_\tau$ which, together, can provide a means of
selecting the most appropriate of the interactions used in EOS
calculations. For example, this combination of values for K$_\infty$
and K$_\tau$ essentially rules out a vast majority of the
Skyrme-type interactions currently in use in nuclear structure
calculations \cite{hs}. This is borne out by Fig.~\ref{fig:kcomp} in
which are plotted the values of $K_{\infty}$ and $K_{\tau}$ for a
number of interactions used in both relativistic and
non-relativistic calculations. It is clear that nearly all of them
fall outside of the ``acceptable'' region defined by the values
obtained in our measurements, leaving a challenge for the theorists
to construct appropriate interactions that meet this criterion.

This work has been supported in part by the U.S. National Science
Foundation (Grants INT03-42942 and PHY04-57120) and by the Japan
Society for Promotion of Science (JSPS).

\end{document}